\documentclass[12pt]{article}

\newcommand{\pslash}{\not \! p}
\newcommand{\kslash}{\not \! k}
\newcommand{\qslash}{\not \! q}

\long\def\symbolfootnote[#1]#2{\begingroup%
\def\thefootnote{\fnsymbol{footnote}}\footnote[#1]{#2}\endgroup}

\usepackage{amsmath,amssymb,amscd,amsbsy,amsgen,amsopn,amstext,
amsxtra}

\usepackage[dvips]{graphicx}

\begin{document}

\begin{flushright}
\end{flushright}

\vskip 0.5 truecm

\begin{center}
{\Large{\bf Lorentz-invariant CPT violation}}
\end{center}
\vskip .5 truecm
\begin{center}
{\bf { Masud Chaichian{$^*$}, Kazuo Fujikawa$^\dagger$ and
Anca Tureanu$^*$}}
\end{center}

\begin{center}
\vspace*{0.4cm} {\it { $^*$Department of Physics, University of Helsinki, P.O.Box 64,\\ FIN-00014 Helsinki,
Finland\\
$^\dagger$ Mathematical Physics Laboratory, RIKEN Nishina Center,\\ Wako 351-0198, Japan}}
\end{center}

\makeatletter
\@addtoreset{equation}{section}
\def\theequation{\thesection.\arabic{equation}}
\makeatother

\begin{abstract}
A Lorentz-invariant CPT violation, which may be termed as long-distance CPT violation in contrast to the familiar short-distance CPT violation, has been recently proposed. This scheme is based on a non-local interaction vertex and characterized by an infrared divergent form factor. We show 
that the Lorentz covariant $T^{\star}$-product is consistently defined and the energy-momentum conservation is preserved in perturbation theory if the path integral is suitably defined for this non-local theory, although unitarity is generally lost. It is illustrated  that T violation is realized in the decay and formation processes. It is also argued that the equality of masses and decay widths of the particle and anti-particle is preserved if the non-local CPT violation is incorporated either directly or as perturbation by starting  with the conventional CPT-even  local Lagrangian. However, we also explicitly show that the present non-local scheme can induce the splitting of particle and anti-particle mass eigenvalues if one considers a more general class of Lagrangians.  
\end{abstract}

\section{Introduction}
The local field theory defined in Minkowski space-time is very successful, and CPT symmetry is a fundamental symmetry of any such  theory~\cite{pauli}. Nevertheless, the possible breaking of CPT symmetry has also been discussed. One of the logical ways to break CPT symmetry is to make the theory non-local by preserving Lorentz symmetry, while the other is to break Lorentz symmetry itself. The Lorentz symmetry breaking scheme 
has been mainly studied in the past including its physical implications~\cite{ellis, greenberg, piguet}.
It is important  to study the possible violation of CPT symmetry in the framework of Lorentz-invariant theory.
A Lorentz-invariant CPT violation, which may be termed as long-distance CPT violation in contrast to the familiar short-distance CPT violation~\cite{ellis}, has been recently proposed  in Ref.~\cite{chaichian}. Its logical consisitency has also been emphasized in~\cite{JGB}. This scheme is based on a non-local interaction vertex and characterized by an infrared-divergent form factor.  
  
To be definite, we study a specific realization of the idea with the Yukawa-type interaction
\begin{eqnarray}
{\cal L}&=&\bar{\psi}(x)[i\gamma^{\mu}\partial_{\mu}-M]\psi(x) +
\frac{1}{2}\partial_{\mu}\phi(x)\partial^{\mu}\phi(x)-\frac{1}{2}m^{2}\phi(x)^{2}\nonumber\\
&&+ g\bar{\psi}(x)\psi(x)\phi(x)- V(\phi)\nonumber\\
&&+ g_{1}\bar{\psi}(x)\psi(x)\int d^{4}y\theta(x^{0}-y^{0})\delta((x-y)^{2}-l^{2})\phi(y)
\end{eqnarray}
as a main theoretical model. This Lagrangian is hermitian and the term with a small real $g_{1}$ and the step function $\theta(x^{0}-y^{0})$ stands for the CTP- and T-violating interaction; $l$ is a real constant parameter. It is interesting that the CPT- and T-violating term is real in the present case. T-violating terms  usually carry imaginary coupling constants in the ordinary local field theory. As for the potential for the scalar field
in (1.1), we have 
\begin{eqnarray}
V(\phi)=\frac{\lambda}{4!}\phi^{4}
\end{eqnarray}
in mind. The Yukawa-type interaction in (1.1) can in fact induce terms odd in $\phi$ as higher-order corrections. We simply eliminate
these corrections by a fine tuning, since the essence of the present  analysis in lower order perturbation theory is not influenced by this procedure. We shall discuss this and related issues in Section 6.
We define the interaction part
\begin{eqnarray}
{\cal L}_{I}&=&g\bar{\psi}(x)\psi(x)\phi(x)\nonumber\\
&&+ g_{1}\bar{\psi}(x)\psi(x)\int d^{4}y\theta(x^{0}-y^{0})\delta((x-y)^{2}-l^{2})\phi(y).
\end{eqnarray}
We treat this non-local Lagrangian in a formal way in path integral as described in~\cite{fujikawa}.

We first note that the present way to introduce CPT violation is based on the extra form factor in momentum space as 
\begin{eqnarray}
&&g_{1}\int d^{4}x\bar{\psi}(x)\psi(x)\int d^{4}y\theta(x^{0}-y^{0})\delta((x-y)^{2}-l^{2})\phi(y)\nonumber\\
&&=g_{1}\int dp_{1}dp_{2}dq\int d^{4}x\bar{\psi}(p_{1})e^{-ip_{1}x}\psi(p_{2})e^{-ip_{2}x}\int d^{4}y\theta(x^{0}-y^{0})\delta((x-y)^{2}-l^{2})\phi(q)e^{-iqy}\nonumber\\
&&=g_{1}\int dp_{1}dp_{2}dq (2\pi)^{4}\delta^{4}(p_{1}+p_{2}+q)\int d^{4}z\theta(z^{0})\delta(z^{2}-l^{2})e^{iqz}\bar{\psi}(p_{1})\psi(p_{2})\phi(q)\nonumber\\
&&=g_{1}\int dp_{1}dp_{2}dq (2\pi)^{4}\delta^{4}(p_{1}+p_{2}+q)\bar{\psi}(p_{1})\psi(p_{2})f(q)\phi(q),
\end{eqnarray}
where we defined 
\begin{eqnarray}
f(q)\equiv \int d^{4}z\theta(z^{0})\delta(z^{2}-l^{2})e^{iqz}.
\end{eqnarray}
Namely, CPT violation is realized by the insertion of the form factor $f(q)$ to the $\phi-\bar\psi{\psi}$ coupling in momentum space. The ordinary local field theory is characterized by $\delta^4(z)$ and $f(q)=1$.  The above form factor is infrared divergent, and it diverges quadratically in the present example. This infrared divergence arises from the fact that we cannot divide Minkowski space into (time-like) domains with finite 
{\em 4-dimensional} volumes in a Lorentz-invariant manner. The Minkowski space is hyperbolic rather than elliptic. CPT symmetry is related to the fundamental structure of Minkowski space, and thus it is gratifying that its possible breaking is also related to the basic properties of Minkowski space.
The details of the infrared divergence of the form factor are discussed in Appendix A.

The above model of CPT violation in \cite{chaichian} was introduced as a counter example to Greenberg's claim~\cite{greenberg} that CPT violation implies Lorentz symmetry violation. This paper is based on different sets of assumptions (see \cite{JGB} for a detailed discussion of those assumptions). To be more specific, Greenberg assumes a consistent axiomatic (Wightman) formulation of realistic particle theory, although to our knowledge nobody has given a convincing axiomatic formulation of the Standard Model, for example. He then assumes that a consistent axiomatic formulation is maintained when one breaks the CPT symmetry simultaneously with Lorentz symmetry, to argue that CPT breaking inevitably breaks Lorentz symmetry. This latter assumption has no solid support, since  a concrete demonstration is missing how the Lorentz invariant theory is inevitably deformed to a Lorentz non-invariant theory, which is also defined in an axiomatic formulation. In this respect, a very de!
 tailed analysis of modified QED, which breaks both Lorentz and CPT symmetries, has been recently given in Ref.~\cite{piguet}. They show that unitarity is generally spoiled unless the charge conjugation symmetry C is preserved. This shows that unitarity is not maintained in full generality with the Lorentz symmetry breaking scheme, since we know that C
invariance is broken in the most realistic models for particles interactions, i.e.  in
the Standard Model and its modifications. Also, the remnant C symmetry would imply that the most interesting aspect of CPT breaking, namely, the particle and antiparticle mass splitting is not realized at least in their model. We find their analysis very interesting and illuminating. Futher analyses in the direction of~\cite{piguet} will greatly enrich our understanding of the  CPT violation with Lorentz symmetry breaking.

 On the other hand, our concrete models break CPT symmetry in a Lorentz-invariant manner, although they inevitably become non-local in space-time. It is thus generally expected that our models as they stand break unitarity. However, it is our opinion that the explicit analyses of the implications of possible CPT breaking, in particular in a Lorentz-invariant scheme, are important. We believe that it is sensible to understand first what the CPT breaking with Lorentz symmetry is by postponing the basic issue of unitarity to future study~\footnote{To our knowledge, there exists no example of a viable field theoretical model of elementary particles which breaks CPT invariance (in addition to C and CP breaking) while preserving unitarity, regardless of whether Lorentz invariance is violated or not.
Therefore, 
it is natural 
to consider the Lorentz invariant CPT violating theories, in
particular the ones studied
in the present work,  as effective theories. The latter could  emerge from a more
fundamental theory  in higher dimensions  or as the low energy limit  of some   basic
theory.}.  Our paper is conceived in this spirit.

\section{Covariant $T^{\star}$-product and energy-momentum conservation}

The non-local field theory has a long history and its general features are well summarized by Marnelius~\cite{marnelius}. It is well known that  theory non-local in time lacks the reliable definition of canonical momenta  and thus canonical quantization. To analyze the aspects of quantized field theory, however, one needs to employ some ways to quantize non-local theories. By employing a procedure close to the canonical quantization, Marnelius shows the breakdown of energy-momentum conservation in this class of theories in addition to the failure in defining a unitary $S$-matrix in general. This shows that one cannot define a consistent canonically quantized energy-momentum operator.

In the present paper, we suggest to employ the path integral which is based on Schwinger's action principle, as explained in detail below. This path integral is based on the equations of motion and an assembly of all sets of Green's functions, which are evaluated in the path integral by one way or another, are identified with quantum field theory. For local field theory, those Green's functions agree with quantum mechanical Green's functions and the canonical structure is extracted from those Green's functions by means of Bjorken--Johnson--Low prescription, which is related to the Riemann--Lebesgue lemma~\cite{fujikawa}. For non-local theory one can still define Green's functions but their direct connection with quantum mechanical Green's functions is not established, although one can show the agreement of those Green's functions with quantum mechanical ones if one considers the local limit of the Lagrangian in the path integral. 
The path integral thus defined is manifestly Poincar\'e invariant for 
Lorentz invariant non-local theoies we study, and thus the energy-momentum  
 conservation is ensured although the canonically quantized energy-momentum operator does not exist. This scheme also provides a covenient scheme to define perturbation theory in terms of the small CPT breaking terms. Ultimately, the test of this scheme depends on the final outcomes of the formulation, namely, whether the final results are physically sensible.

We thus start with the path integral by integrating the formal  equations of motion by means of Schwinger's action principle~\cite{fujikawa}, whose basis is analogous to that of the  Yang--Feldman formulation~\cite{yang}. Namely,
\begin{eqnarray}
\langle 0,+\infty|0,-\infty\rangle_{J}=\int{\cal D}\bar{\psi}{\cal D}\psi{\cal D}\phi\exp\left\{i\int d^{4}x [
{\cal L}_{0}+{\cal L}_{I}+{\cal L}_{J}]\right\}
\end{eqnarray}
with the source term ${\cal L}_{J}=\bar{\psi}(x)\eta(x)+\bar{\eta}(x)\psi(x)+\phi(x)J(x)$, 
and one may  generate Green's functions in a power-series expansion of perturbation as
\begin{eqnarray}
(i)^{n}\langle T^{\star}\phi(x_{1})...\phi(x_{N})\int d^{4}y_{1}{\cal L}_{I}(y_{1})
....\int d^{4}y_{n}{\cal L}_{I}(y_{n})\rangle,
\end{eqnarray}
where we consider only $N$ scalar particles corresponding to external fields, for simplicity. We use the $T^{\star}$-product which is essential to make the path integral on the basis of Schwinger's action principle consistent~\cite{fujikawa}.

This scheme of path integral may be regarded as a definition of quantization in the present paper and it was successfully applied to the analysis of theories in non-commutative space-time~\cite{fujikawa}. It was shown there that one can reproduce all the results of canonical quantization rules for higher-derivative theory by using the Bjorken--Johnson--Low prescription in conventional flat space-time, and that one can reproduce all the results of Yang--Feldman formulation at least in a perturbative sense in non-commutative space-time. Pauli's spin-statistics theorem
is also tightly related to the CPT symmetry, although for the former theorem, one needs
stronger assumptions (see the first paper in \cite{chaichian2}). In this respect, we mention that the spin-statistics theorem can be defined in the path integral framework~\cite{fujikawa2}. In particular, the statistics of fields  appearing in the free part of the Yukawa theory, which defines perturbation theory, can be readily handled by the path integral. The path integral on the basis of Schwinger's action principle is thus expected to provide a reasonable framework to deal with the present non-local theory, although a scheme satisfactory in all respects is absent.  Our analysis in this paper is based on this path integral quantization and formal perturbative expansion.
   
The part of the above Green's function which depends on the CPT-violating interaction is written as 
\begin{eqnarray}
&&(ig_{1})^{n}\int d^{4}y_{1}\int d^{4}z_{1}....\int d^{4}y_{n}\int d^{4}z_{n}\nonumber\\
&&\times\theta(y_{1}^{0}-z_{1}^{0})\delta((y_{1}-z_{1})^{2}-l^{2})...\theta(y_{n}^{0}-z_{n}^{0})\delta((y_{n}-z_{n})^{2}-l^{2})\nonumber\\
&&\times\langle T^{\star}\phi(x_{1})...\phi(x_{N})\bar{\psi}(y_{1})\psi(y_{1})\phi(z_{1})....\bar{\psi}(y_{n})\psi(y_{n})\phi(z_{n})\rangle.
\end{eqnarray}
This expression is manifestly Lorentz invariant in perturbation theory, since the correlation function of fields is defined in terms of free fields of Yukawa theory and the extra constraints imposed on the correlation function are Lorentz invariant.

If one defines the correlation function by
\begin{eqnarray}
&&G(x_{1},...,x_{N}; y_{1},z_{1}, ......,y_{n},z_{n})\nonumber\\
&&=\langle T^{\star}\phi(x_{1})...\phi(x_{N})\bar{\psi}(y_{1})\psi(y_{1})\phi(z_{1})....\bar{\psi}(y_{n})\psi(y_{n})\phi(z_{n})\rangle,
\end{eqnarray}
which is expressible in terms of ordinary free propagators of Yukawa theory, the expression is 
invariant under the uniform constant shift of coordinates
\begin{eqnarray}
&&G(x_{1}+\epsilon,...,x_{N}+\epsilon; y_{1}+\epsilon,z_{1}+\epsilon, ......,y_{n}+\epsilon,z_{n}+\epsilon)\nonumber\\
&&=G(x_{1},...,x_{N}; y_{1},z_{1}, ......,y_{n},z_{n}),
\end{eqnarray}
where $\epsilon$ is a constant 4-vector.
 This translational symmetry, which also keeps the factor 
\begin{eqnarray}
&&\int d^{4}y_{1}\int d^{4}z_{1}....\int d^{4}y_{n}\int d^{4}z_{n}\nonumber\\
&&\times\theta(y_{1}^{0}-z_{1}^{0})\delta((y_{1}-z_{1})^{2}-l^{2})...\theta(y_{n}^{0}-z_{n}^{0})\delta((y_{n}-z_{n})^{2}-l^{2})
\end{eqnarray}
invariant, ensures the energy-momentum conservation in the sense of the aboved perturbation theory (2.3), although we do not have the quantized energy-momentum operator in the conventional sense. Namely, the in-coming and out-going 4-momenta carried by the external fields $\phi(x_{1}),...,\phi(x_{N})$ are equal.

 The operator statement of energy-momentum conservation is ill-defined in non-local Lagrangians in general, as was already explained. However, the action defined by our Lagrangian (1.1), as it stands, is local if one looks at the fermion and the boson separately, and thus it is much better defined than general non-local theories. In fact, the form factor (1.5) is  close to the Coulomb potential for the space-like momentum $q$, as is shown in Appendix A, and in this sense our Lagrangian is similar to a system interacting through the non-local Coulomb potential. Thus the formal canonical
energy-momentum operator, if one defines it for the present model, may have a better property. 

As for the absence of a covariant time-ordered product,
the Lorentz-invariant $T^{\star}$-product can be defined as above
in a perturbative sense. But the absence of the unitary $S$-matrix is still there in our formulation since the cutting rule to analyze perturbative unitarity is not precisely defined for the amplitude in (2.3) involving interaction terms non-local in time. This difficulty in fact persists in any known scheme.

\section{T violation in decay and production processes}

We start with a comment on a novel feature of the present scheme of CPT and T violation. The part of the interaction 
\begin{eqnarray}
{\cal L}_{I}^{\prime}=g_{1}\bar{\psi}(x)\psi(x)\int d^{4}y\theta(x^{0}-y^{0})\delta((x-y)^{2}-l^{2})\phi(y)
\end{eqnarray}
clearly violates T symmetry in a quantum mechanical sense. But in a field theoretical sense, it has the following special properties.
When one considers the limit $x^{0}\rightarrow -\infty$, the above interaction goes to zero:
\begin{eqnarray}
{\cal L}_{I}^{\prime}=g_{1}\bar{\psi}(x)\psi(x)\int d^{4}y\theta(x^{0}-y^{0})\delta((x-y)^{2}-l^{2})\phi(y)\rightarrow 0,
\end{eqnarray}
regardless of the behavior of the coupling constant $g_{1}$. Also, for
$x^{0}\rightarrow \infty$, we have
\begin{eqnarray}
{\cal L}_{I}^{\prime}&=&g_{1}\bar{\psi}(x)\psi(x)\int d^{4}y\theta(x^{0}-y^{0})\delta((x-y)^{2}-l^{2})\phi(y)\nonumber\\
&\rightarrow& g_{1}\bar{\psi}(x)\psi(x)\int d^{4}y\delta((x-y)^{2}-l^{2})\phi(y),
\end{eqnarray}
regardless of the behavior of the coupling constant $g_{1}$. 
Namely, the effect of the T violation disappears independently of the behavior of the coupling constant $g_{1}$ in asymptotic regions.

It is thus desirable to confirm that the present way to introduce CPT and T violation really leads to T violation in the conventional sense. For this purpose, we analyze the decay  $\phi \rightarrow \bar{\psi}\psi$ and its inverse, the production process $\bar{\psi}\psi \rightarrow \phi$.

\subsection { Tree processes}

We start with the analysis of tree diagrams.
We first evaluate the decay process $\phi\rightarrow \bar{\psi}\psi$
by assuming $m > 2M$:
\begin{eqnarray}
&&\langle 0|a(\vec{p}_{1})b(\vec{p}_{2})i\int d^{4}x{\cal L}_{I}(x)c^{\dagger}(\vec{k})|0\rangle\nonumber\\
&&=ig\int d^{4}x e^{ip_{1}x}e^{ip_{2}x}e^{-ikx}+ig_{1}\int d^{4}x\int d^{4}y\theta(x^{0}-y^{0})\delta((x-y)^{2}-l^{2})e^{ip_{1}x}e^{ip_{2}x}e^{-iky}\nonumber\\
&&=(2\pi)^{4}\delta^{4}(p_{1}+p_{2}-k)ig+(2\pi)^{4}\delta^{4}(p_{1}+p_{2}-k)ig_{1}\int d^{4}z\theta(z^{0})\delta(z^{2}-l^{2})e^{ikz}\nonumber\\
&&=(2\pi)^{4}\delta^{4}(p_{1}+p_{2}-k)i\left[g
+g_{1}\int d^{4}z\theta(z^{0})\delta(z^{2}-l^{2})e^{ikz}\right],
\end{eqnarray}
where we used the operator notation, but the same result is obtained by starting with Green's function and amputating external legs since the free part of Yukawa theory can be equally formulated in the  path integral and operator formulation. Some irrelevant constant factors were omitted. We also suppressed the spinor index and left aside the factor  $\bar{u}(\vec{p}_{1})v(\vec{p}_{2})$ by which  this amplitude should actually be multiplied.

This may be compared to the time-reversed pair annihilation process $\bar{\psi}\psi\rightarrow \phi$:
\begin{eqnarray}
&&\langle 0|c(-\vec{k})i\int d^{4}x{\cal L}_{I}(x)a^{\dagger}(-\vec{p}_{1})b^{\dagger}(-\vec{p}_{2})|0\rangle\nonumber\\
&&=ig\int d^{4}x e^{-ip_{1}x}e^{-ip_{2}x}e^{ikx}+ig_{1}\int d^{4}x\int d^{4}y\theta(x^{0}-y^{0})\delta((x-y)^{2}-l^{2})e^{-ip_{1}x}e^{-ip_{2}x}e^{iky}\nonumber\\
&&=(2\pi)^{4}\delta^{4}(p_{1}+p_{2}-k)ig+(2\pi)^{4}\delta^{4}(p_{1}+p_{2}-k)ig_{1}\int d^{4}z\theta(z^{0})\delta(z^{2}-l^{2})e^{-ikz}\nonumber\\
&&=(2\pi)^{4}\delta^{4}(p_{1}+p_{2}-k)i\left[g
+g_{1}\int d^{4}z\theta(z^{0})\delta(z^{2}-l^{2})e^{-ikz}\right].
\end{eqnarray}
This amplitude should be multiplied by $\bar{u}(-\vec{p}_{1})v(-\vec{p}_{2})$ to express the time reversed process with reversed momentum directions.

We recognize that these two amplitudes contain different phases due to the T-violating term. This T violation is characterized by the form factors
\begin{eqnarray}
&&f_{\pm}(k)=\int d^{4}z_{1}
e^{\pm ikz_{1}}\theta(z_{1}^{0})\delta((z_{1})^{2}-l^{2}),
\end{eqnarray}
which are  inequivalent for time-like $k$ due to the factor $\theta(z_{1}^{0})$.
For the time-like momentum  $k$ one may choose a suitable Lorentz frame such that $\vec{k}=0$, and 
\begin{eqnarray}
f_{\pm}(k^{0})
&=&2\pi \int_{0}^{\infty}dz\frac{z^{2}e^{\pm ik^{0}\sqrt{z^{2}+l^{2}}}}{\sqrt{z^{2}+l^{2}}},
\end{eqnarray}
while for the space-like momentum  $k$ one may choose a suitable Lorentz frame such that $k^{0}=0$, and
\begin{eqnarray}
f_{\pm}(\vec{k}) 
&=&\frac{2\pi}{|k|^{2}}\int_{0}^{\infty} dz z \frac{\sin z}{\sqrt{z^{2}+(|k|l)^{2}}},
\end{eqnarray}
which is analogous to the Fourier transform of the Coulomb potential and real. The details of these calculations are found in Appendix A.
It is also explained there that $f_{\pm}(k)$ is mathematically related to the formula of the two-point Wightman function (for a free scalar field), which suggests that 
$f_{\pm}(k)$ is mathematically well-defined for $k\neq 0$ at least in the sense of distribution.

In our application, we have time-like momentum with $k^{0}=m>0$, and thus 
\begin{eqnarray}
f_{\pm}(m)
&=&\frac{2\pi}{m^{2}} \int_{0}^{\infty}dz\frac{z^{2}e^{\pm i\sqrt{z^{2}+(ml)^{2}}}}{\sqrt{z^{2}+(ml)^{2}}}
\end{eqnarray}
is the relevant phase factor. This quantity can be made real for $ml\rightarrow 0$ by a suitable rotation of the integration contour in the complex $z$-plane. But for $ml \neq 0$, one cannot eliminate the phases arising from the approximate interval  $[0, \pm iml]$ of $z$ by a rotation of the integration contour. We thus have non-vanishing T-violating phases. But one cannot detect the T violation in the absolute square of the above probability amplitudes in (3.4) and (3.5) at this tree level.

\subsection { One-loop correction}

A way to recognize the effects of T violation in the present scheme may be to include one-loop corrections to the decay and production processes. 
The basic idea is that the decay amplitude develops its own intrinsic
imaginary parts in the one-loop corrections if one chooses the masses to be   
\begin{eqnarray}
3M> m > 2M,
\end{eqnarray}
and we can detect the T-violating phase by looking at the interference of two complex quantities. The specific choice (3.10) makes the present decay mode the only allowed decay mode and thus the analysis becomes more definite. We consider only the lowest order effect in $g_{1}$ by assuming
that $g_{1}$ is substantially smaller than $g$, and the perturbation theory is still meanigful. Namely,
\begin{eqnarray}
1 \gg g \gg g_{1}. 
\end{eqnarray}

We now analyze the diagrams with one-loop corrections. To this order, we have amplitudes with the coupling constants:
\begin{eqnarray}
&&1 \gg O(g) \gg O(g_{1}),\nonumber\\
&&O(g) \gg O(g^{3}) \gg O(g_{1}g^{2}),\nonumber\\
 &&O(g_{1}) \gg O(g_{1}g^{2}). 
\end{eqnarray}
We thus retain only the leading diagrams of the order $O(g)$, $O(g^{3})$ and 
$O(g_{1})$. In this way, we can reduce the number of Feynman diagrams 
substantially.

We need to evaluate only the one-loop diagrams with $O(g^{3})$, which are the one-loop corrections in the ordinary Yukawa theory described by (1.1).
We have three kinds of diagrams. The vertex correction for $\phi-\psi\bar{\psi}$ coupling, the self-energy correction to the scalar $\phi$, and the self-energy correction for the fermion $\psi$.
Among these three kinds of diagrams, the self-energy correction to the fermion does not develop an imaginary part for
\begin{eqnarray}
m+M > M ,
\end{eqnarray}
since we are assuming $m > 2M$, and thus the fermion $\psi$ cannot decay to $\psi+\phi$. The tree amplitude with $O(g)$ dominates the real part of the amplitude, and we can neglect the real part contribution from the fermion self-energy correction.

We thus need to evaluate only the $O(g^{3})$ amplitudes of  the vertex $\phi-\psi\bar{\psi}$ and the self-energy of the scalar $\phi$ in the ordinary Yukawa interaction. The one-loop correction to the vertex of the decay process  in Yukawa theory has the standard form,
\begin{eqnarray}
&&(ig)^{3}(2\pi)^{4}\delta^{4}(p_{1}+p_{2}-k)\int\frac{d^{4}q}{(2\pi)^{4}}\nonumber\\
&&\times\bar{u}(p_{1})\left[\frac{i}{\qslash+\kslash-M+i\epsilon}\frac{i}{\qslash-\pslash_{2}-M+i\epsilon}\frac{i}{q^{2}-m^{2}+i\epsilon}\right]v(p_{2}).
\end{eqnarray}
This integral is standard and evaluated by various methods in field theory. One needs to take out one of the imaginary factor $i$ from this amplitude when one talks about real and imaginary parts in conformity with (3.4) and (3.5); actually, another factor of $i$ appears due to the Wick rotation of the loop momentum $q$.  The following facts about this amplitude are known: the real part contains a logarithmic divergence which is removed by the coupling constant renormalization. The imaginary part, which is controlled by the Landau--Cutkosky rule, is finite. After using the Dirac equation, the final result of the renormalized vertex
depends only on Lorentz-invariant combinations such as $p_{1}\cdot k$, which are invariant under the reversal of all the momentum directions, and thus {\em common} to decay and formation processes.   
The evaluation of the self-energy correction of the scalar $\phi$ proceeds in a similar manner.

After these calculations of Feynman diagrams, we obtain in a symbolic notation:
\begin{eqnarray}
A(\phi \rightarrow \bar{\psi}\psi)&=& i\left[g|A_{1}|+ g^{3}|A_{3}|e^{-i\theta_{i}}+ g_{1}|A^{\prime}_{1}|e^{i\theta_{CPT}}\right]\bar{u}(\vec{p}_{1})v(\vec{p}_{2}),\nonumber\\
A(\bar{\psi}\psi \rightarrow \phi)&=& i\left[g|A_{1}|+ g^{3}|A_{3}|e^{-i\theta_{i}}+ g_{1}|A^{\prime}_{1}|e^{-i\theta_{CPT}}\right]\bar{u}(-\vec{p}_{1})v(-\vec{p}_{2}),
\end{eqnarray}
where $\theta_{i}$ stands for the intrinsic imaginary phase of the decay or formation amplitudes in Yukawa theory and $\theta_{CPT}$ stands for the phase of the CPT-violating term. The intrinsic phase is the same for the two reversed processes, as we analyzed above. They arise from  essentially the same Feynman amplitudes in  T-invariant Yukawa theory. We have already evaluated the tree amplitudes $g|A_{1}|+ g_{1}|A^{\prime}_{1}|e^{\pm i\theta_{CPT}}$ in (3.4) and (3.5).

We can thus produce the time-reversal non-invariance in the square of the  probability amplitudes,
\begin{eqnarray}
|A(\phi \rightarrow \bar{\psi}\psi)|^{2} \neq |A(\bar{\psi}\psi \rightarrow \phi)|^{2},
\end{eqnarray}
after averaging over spin directions for the processes $\phi \rightarrow \bar{\psi}\psi$ and $\bar{\psi}\psi \rightarrow \phi$, as a result of the interference of the two phases, $\theta_{i}$ and $\pm\theta_{CPT}$. 
 
Our analysis shows that the present CPT- and T-violation
scheme, which may be termed as CPT violation at long distances in 
contrast to the more familiar CPT violation at short distances, produces the expected T violation  in the path integral quantization. 

\subsection{Formal analysis}

We here briefly mention a formal analysis which supports our formula 
(3.15). In the analysis of T symmetry defined by the anti-unitary ${\cal T} $ operator, it is convenient to consider the 
quantity
\begin{eqnarray}
&&\langle A(t_{f}), \exp[-iH(t_{f}-t_{i})]B(t_{i})\rangle\nonumber\\
&&=\langle A(t_{f}),{\cal T}^{-1}{\cal T} \exp[-iH(t_{f}-t_{i})]{\cal T}^{-1}{\cal T}B(t_{i})\rangle\nonumber\\
&&=\langle {\cal T}A(t_{f}),{\cal T} \exp[-iH(t_{f}-t_{i})]{\cal T}^{-1}{\cal T}B(t_{i})\rangle^{\star}\nonumber\\
&&=\langle {\cal T}A(t_{f}),\exp[i{\cal T} H{\cal T}^{-1}(t_{f}-t_{i})]{\cal T}B(t_{i})\rangle^{\star}\nonumber\\
&&=\langle \tilde{B}(-t_{i}),\exp\{-i({\cal T} H{\cal T}^{-1})^{\dagger}[(-t_{i})-(-t_{f})]\}\tilde{A}(-t_{f})\rangle,
\end{eqnarray}
where $\tilde{A}(-t_{f})={\cal T}A(t_{f})$ and $\tilde{B}(-t_{i})={\cal T}B(t_{i})$ are the time-reversed states of $A(t_{f})$ and $B(t_{i})$, respectively. When one considers the limit of $t_{f}\rightarrow \infty$ and $t_{i}\rightarrow
-\infty$, respectively, the above relation gives the $U$-matrix relation for the normal and "time-reversed" processes, although the "time-reversed" process is generated by $({\cal T} H{\cal T}^{-1})^{\dagger}$. 

Now one can {\em directly} evaluate the time-reversed process for a given Hamiltonian $H$, which may or may not be time-reversal invariant, 
as 
\begin{eqnarray}
\langle \tilde{B}(-t_{i}),\exp\{-iH[(-t_{i})-(-t_{f})]\}\tilde{A}(-t_{f})\rangle.
\end{eqnarray}
The agreement of this expression with the last expression in (3.17) is the condition of time reversal invariance. 
We usually choose the Hamiltonian to be hermitian, and thus the time reversal invariance is equivalent to 
\begin{eqnarray}
{\cal T} H{\cal T}^{-1}=H.
\end{eqnarray}

In our application, we have a Hamiltonian of the form
\begin{eqnarray}
H=H_{Yukawa}+H_{CPT},
\end{eqnarray}
where $H_{Yukawa}$ stands for the pure Yukawa theory. This $H$ satisfies
\begin{eqnarray}
{\cal T}H{\cal T}^{-1}&=&
H_{Yukawa}+H^{\prime}_{CPT},
\end{eqnarray}
with
\begin{eqnarray}
H^{\prime}_{CPT}={\cal T}H_{CPT}{\cal T}^{-1}\neq H_{CPT}.
\end{eqnarray}
Here, $H_{CPT}$ stands for the time-reversal violating part 
\begin{eqnarray}
H_{CPT}=-g_{1}\int d^{3}x \bar{\psi}(x)\psi(x)\int d^{4}y\theta(x^{0}-y^{0})\delta((x-y)^{2}-l^{2})\phi(y),
\end{eqnarray}
formally treated as a lowest-order perturbation in our spirit of path integral quantization.

In  the analysis of time-reversal invariance, one compares
\begin{eqnarray}
\langle A(t_{f}), \exp[-iH(t_{f}-t_{i})]B(t_{i})\rangle
\end{eqnarray}
with the directly evaluated reversed process
\begin{eqnarray}
\langle \tilde{B}(-t_{i}),\exp\{-iH[(-t_{i})-(-t_{f})]\}\tilde{A}(-t_{f})\rangle,
\end{eqnarray}
or equivalently, 
\begin{eqnarray}
\langle \tilde{B}(-t_{i}),\exp\{-i(H_{Yukawa}+H^{\prime}_{CPT})[(-t_{i})-(-t_{f})]\}\tilde{A}(-t_{f})\rangle,
\end{eqnarray}
with
\begin{eqnarray}
\langle \tilde{B}(-t_{i}),\exp\{-i(H_{Yukawa}+H_{CPT})[(-t_{i})-(-t_{f})]\}\tilde{A}(-t_{f})\rangle.
\end{eqnarray}
In these expressions, $H^{\prime}_{CPT}$ and $H_{CPT}$ generate
different phases.

If one considers the processes lowest order in $H_{CPT}$ or $H^{\prime}_{CPT}$, the real and imaginary parts generated by $H_{Yukawa}$  in perturbation theory in (3.26) and (3.27) are identical for those two amplitudes; this agrees with our one-loop Feynman diagram analysis. The vertex correction, for exmple, is a counterpart of the phase shift induced by strong interactions. One can thus observe the interference of those two amplitudes generated by $H_{CPT}$ and $H_{Yukawa}$ (or $H^{\prime}_{CPT}$ and  $H_{Yukawa}$) to detect the effects of the time reversal non-invariance of $H_{CPT}$,  
\begin{eqnarray}
&&|\langle \tilde{B}(-t_{i}),\exp\{-i(H_{Yukawa}+H^{\prime}_{CPT})[(-t_{i})-(-t_{f})]\}\tilde{A}(-t_{f})\rangle|^{2}\nonumber\\
&&\neq |\langle \tilde{B}(-t_{i}),\exp\{-i(H_{Yukawa}+H_{CPT})[(-t_{i})-(-t_{f})]\}\tilde{A}(-t_{f})\rangle|^{2},
\end{eqnarray}
or equivalently
\begin{eqnarray}
&&|\langle A(t_{f}), \exp[-i(H_{Yukawa}+H_{CPT})(t_{f}-t_{i})]B(t_{i})\rangle|^{2}\nonumber\\
&&\neq |\langle \tilde{B}(-t_{i}),\exp\{-i(H_{Yukawa}+H_{CPT})[(-t_{i})-(-t_{f})]\}\tilde{A}(-t_{f})\rangle|^{2},
\end{eqnarray}
in the limit $t_{f}\rightarrow \infty$ and $t_{i}\rightarrow
-\infty$.

To detect the interference, we need two separate channels as in the double-slit experiment: one is generated by $H_{Yukawa}$ and the other by $H_{CPT}$ in our example.

\section{Equality of mass and width of the particle and anti-particle}

To examine the effects of CPT violation on the equality of masses and widths of fermion and anti-fermion in our model, we start with some sample calculations in perturbation theory by incorporating the CPT-violating interaction. We show that our calculation leads to reasonable results which suggest that our path integral prescription is sensible, but no mass or width splitting emerges, since the models we analyze contain C or CP as residual symmetry. Nevertheless, our calculations clearly illustrate that the formal hermiticity  of the Lagrangian tends to mask the possible effects of CPT violation on masses and widths in the present scheme.

\subsection{ One-loop fermion self-energy corrections}

The fermion propagator is given by
\begin{eqnarray}
\langle T^{\star}\psi(w_{1})\bar{\psi}(w_{2})\rangle
\end{eqnarray}
and the second-order perturbation in the T-violating interaction gives rise to 
\begin{eqnarray}
&&(ig_{1})^{2}\int d^{4}xd^{4}z_{1} d^{4}yd^{4}z_{2}
S_{F}(w_{1}-x)\theta(x^{0}-z_{1}^{0})\delta((x-z_{1})^{2}-l^{2})
\nonumber\\
&\times& S_{F}(x-y)D_{F}(z_{1}-z_{2})\theta(y^{0}-z_{2}^{0})\delta((y-z_{2})^{2}-l^{2})S_{F}(y-w_{2})\nonumber\\
&=&(ig_{1})^{2}\int d^{4}xd^{4}z_{1} d^{4}yd^{4}z_{2}
S_{F}(w_{1}-x)\theta(z_{1}^{0})\delta((z_{1})^{2}-l^{2})
\nonumber\\
&\times& S_{F}(x-y)D_{F}(x-y-z_{1}+z_{2})\theta(z_{2}^{0})\delta((z_{2})^{2}-l^{2})S_{F}(y-w_{2}).
\end{eqnarray}
After truncating the external legs, this leads to the self-energy correction, 
\begin{eqnarray}
\Sigma(p)&=&(g_{1})^{2}\int\frac{d^{4}q}{(2\pi)^{4}}\frac{d^{4}k}{(2\pi)^{4}}\int d^{4}xd^{4}z_{1} d^{4}yd^{4}z_{2}
e^{-ip^{\prime}(-x)}\theta(z_{1}^{0})\delta((z_{1})^{2}-l^{2})
\nonumber\\
&\times& \frac{e^{-iq(x-y)}}{\qslash -m}\frac{e^{-ik(x-y-z_{1}+z_{2})}}{k^{2}-m^{2}}\theta(z_{2}^{0})\delta((z_{2})^{2}-l^{2})e^{-ipy}\nonumber\\
&=&(g_{1})^{2}(2\pi)^{4}\delta^{(4)}(p^{\prime}-p)\int\frac{d^{4}k}{(2\pi)^{4}}
\frac{1}{\pslash-\kslash -m}\frac{1}{k^{2}-m^{2}}F(k^{2}),
\end{eqnarray}
with
\begin{eqnarray}
F(k^{2})&=&\int d^{4}z_{1}d^{4}z_{2}
\theta(z_{1}^{0})\delta((z_{1})^{2}-l^{2})
 e^{ikz_{1}-ikz_{2}}\theta(z_{2}^{0})\delta((z_{2})^{2}-l^{2})\nonumber\\
 &=&f_{+}(k)f_{-}(k),
\end{eqnarray}
where $f_{\pm}(k)$ is defined in (3.6).
This form factor is Lorentz invariant and also invariant under $k\rightarrow-k$, and 
thus it depends only on $k^{2}$. The form factor $F(k^{2})$ cannot induce the splitting of positive and negative $p_{0}$ in (4.4). Namely, no splitting of particle and anti-particle masses.

We next evaluate the corrections arising from the mixing of T-violating and T-preserving interactions:
\begin{eqnarray}
&&i^{2}g_{1}g\int d^{4}xd^{4}z_{1} d^{4}y
S_{F}(w_{1}-x)\theta(x^{0}-z_{1}^{0})\delta((x-z_{1})^{2}-l^{2})
 S_{F}(x-y)D_{F}(z_{1}-y)S_{F}(y-w_{2})\nonumber\\
&+&i^{2}g_{1}g\int d^{4}xd^{4}yd^{4}z_{2}
S_{F}(w_{1}-x) S_{F}(x-y)D_{F}(x-z_{2})\theta(y^{0}-z_{2}^{0})\delta((y-z_{2})^{2}-l^{2})S_{F}(y-w_{2})\nonumber\\
&=&i^{2}g_{1}g\int d^{4}xd^{4}z_{1} d^{4}y
S_{F}(w_{1}-x)\theta(z_{1}^{0})\delta((z_{1})^{2}-l^{2})
 S_{F}(x-y)D_{F}(x-y-z_{1})S_{F}(y-w_{2})\nonumber\\
&+&i^{2}g_{1}g\int d^{4}xd^{4}yd^{4}z_{2}
S_{F}(w_{1}-x)
 S_{F}(x-y)D_{F}(x-y+z_{2})\theta(z_{2}^{0})\delta((z_{2})^{2}-l^{2})S_{F}(y-w_{2}).\nonumber\\
\end{eqnarray}
After truncating the external legs, we obtain the self-energy corrections:
\begin{eqnarray}
	\Sigma(p)&=&g_{1}g\int\frac{d^{4}q}{(2\pi)^{4}}\frac{d^{4}k}{(2\pi)^{4}}\int d^{4}xd^{4}z_{1} d^{4}y
e^{-ip^{\prime}(-x)}\theta(z_{1}^{0})\delta((z_{1})^{2}-l^{2})
 \frac{e^{-iq(x-y)}}{\qslash -m}\frac{e^{-ik(x-y-z_{1})}}{k^{2}-m^{2}}e^{-ipy}\nonumber\\
&+&g_{1}g\int\frac{d^{4}q}{(2\pi)^{4}}\frac{d^{4}k}{(2\pi)^{4}}\int d^{4}xd^{4}yd^{4}z_{2}
e^{-ip^{\prime}(-x)}
\frac{e^{-iq(x-y)}}{\qslash -m}\frac{e^{-ik(x-y+z_{2})}}{k^{2}-m^{2}}\theta(z_{2}^{0})\delta((z_{2})^{2}-l^{2})e^{-ipy}\nonumber\\
&=&g_{1}g(2\pi)^{4}\delta^{(4)}(p^{\prime}-p)\int\frac{d^{4}k}{(2\pi)^{4}}
\frac{1}{\pslash-\kslash -m}\frac{1}{k^{2}-m^{2}}(f_{+}(k)+f_{-}(k)),
\end{eqnarray}
where we used the form factors in (3.6).
 We have the combination of form factors
\begin{eqnarray}
&&f_{+}(k)+f_{-}(k)\nonumber\\
&&=\int d^{4}z[e^{ik_{0}z^{0}}\theta(z^{0})+e^{-ik_{0}z^{0}}\theta(z^{0})]e^{i\vec{k}\vec{z}}\delta((z)^{2}-l^{2})\nonumber\\
&&=\int d^{4}z[e^{ik_{0}z^{0}}\theta(z^{0})+e^{ik_{0}z^{0}}\theta(-z^{0})]e^{i\vec{k}\vec{z}}\delta((z)^{2}-l^{2})\nonumber\\
&&=\int d^{4}z
e^{ikz}\delta((z)^{2}-l^{2})\nonumber\\
&&\equiv f_{0}(k^{2}),
\end{eqnarray}
which contains no T-violating step function.
Namely, the mixing terms give rise to 
\begin{eqnarray}
g_{1}g(2\pi)^{4}\delta^{(4)}(p^{\prime}-p)\int\frac{d^{4}k}{(2\pi)^{4}}
\frac{1}{\pslash-\kslash -m}\frac{1}{k^{2}-m^{2}}f_{0}(k^{2}),
\end{eqnarray}
which shows no splitting between the negative and positive $p_{0}$.

We thus see no indication of particle and anti-particle mass splitting in this setting. This is what we expect since  our initial Lagrangian is invariant under C and CP transformations, which ensure the equality between the masses of the particle and the anti-particle. But an important implication of these calculations is that the symmetric combination of CPT-violating form factors such as $f_{+}(k)f_{-}(k)$  and $f_{+}(k)+f_{-}(k)$ ensures the symmetry between the negative and positive $p_{0}$. This is also regarded as a result of the hermiticity of the amplitudes.

\subsection { Inclusion of CP violation}

The simplest way to incorporate CP violation into the above scheme
is to consider a modification of the  Lagrangian (1.1) as follows:
\begin{eqnarray}
{\cal L}&=&\bar{\psi}(x)[i\gamma^{\mu}\partial_{\mu}-M]\psi(x) +
\frac{1}{2}\partial_{\mu}\phi(x)\partial^{\mu}\phi(x)-\frac{1}{2}m^{2}\phi(x)^{2}\nonumber\\
&&+ g\bar{\psi}(x)(1+i\epsilon\gamma_{5})\psi(x)\phi(x)- V(\phi)\nonumber\\
&&+ g_{1}\bar{\psi}(x)\psi(x)\int d^{4}y\theta(x^{0}-y^{0})\delta((x-y)^{2}-l^{2})\phi(y),
\end{eqnarray}
with a small $g_{1}$ for the CPT- and T-violating interaction, and the small real $\epsilon$ term violating P and T and thus CP; the chiral freedom was fixed by choosing the fermion mass term real.
We define the interaction part
\begin{eqnarray}
{\cal L}_{I}&=&g\bar{\psi}(x)(1+i\epsilon\gamma_{5})\psi(x)\phi(x)\nonumber\\
&+& g_{1}\bar{\psi}(x)\psi(x)\int d^{4}y\theta(x^{0}-y^{0})\delta((x-y)^{2}-l^{2})\phi(y),
\end{eqnarray}
which is treated in the path integral quantization.

We consider the contribution arising from the mixing of two classes of  interactions. The self-energy correction in (4.6) is then replaced by 
\begin{eqnarray}
\Sigma(p)
&=&i^{2}g_{1}g\int d^{4}xd^{4}z_{1} d^{4}y
e^{-ip^{\prime}(-x)}\theta(z_{1}^{0})\delta((z_{1})^{2}-l^{2})
\nonumber\\
&\times& S_{F}(x-y)(1+i\epsilon\gamma_{5})D_{F}(x-y-z_{1})e^{-ipy}\nonumber\\
&+&i^{2}g_{1}g\int d^{4}x d^{4}yd^{4}z_{2}
e^{-ip^{\prime}(-x)}
\nonumber\\
&\times& (1+i\epsilon\gamma_{5})S_{F}(x-y)D_{F}(x-y+z_{2})\theta(z_{2}^{0})\delta((z_{2})^{2}-l^{2})e^{-ipy}\nonumber\\
&=&g_{1}g\int\frac{d^{4}q}{(2\pi)^{4}}\frac{d^{4}k}{(2\pi)^{4}}(2\pi)^{4}\delta^{(4)}(p^{\prime}-q-k)(2\pi)^{(4)}\delta^{4}(p-q-k)
\\
&\times&\left [\frac{1}{\qslash -m}\frac{1}{k^{2}-m^{2}}(1+i\epsilon\gamma_{5})f_{+}(k)+(1+i\epsilon\gamma_{5})\frac{1}{\qslash -m}\frac{1}{k^{2}-m^{2}}f_{-}(k)\right],\nonumber
\end{eqnarray}
which is written as 
\begin{eqnarray}
\Sigma(p)
&=&g_{1}g(2\pi)^{4}\delta^{(4)}(p^{\prime}-p)\int\frac{d^{4}k}{(2\pi)^{4}}\Big\{\frac{1}{\pslash-\kslash -m}\frac{1}{k^{2}-m^{2}}
\left(f_{+}(k)+f_{-}(k)\right)\nonumber\\
&&+i\epsilon\gamma_{5}\left[\left(\frac{1}{-\pslash+\kslash -m}\frac{1}{k^{2}-m^{2}}\right)f_{+}(k)+\frac{1}{\pslash-\kslash -m}\frac{1}{k^{2}-m^{2}}f_{-}(k)\right]\Big\},\nonumber\\
\end{eqnarray}
where we used $f_{+}(k)$ and $f_{-}(k)$ in (3.6), which are invariant under the proper Lorentz transformation, and $f_{+}(-k)=f_{-}(k)$. We thus have the form factor $f_{+}(k)+f_{-}(k)=f_{0}(k^{2})$ for the CP-invariant part of (4.12) as in (4.7),
which is invariant under $k\rightarrow -k$. 

The CP-invariant part contained in the mixing of the T-invariant and T-violating interactions thus gives rise to 
\begin{eqnarray}
&&g_{1}g(2\pi)^{4}\delta^{(4)}(p^{\prime}-p)\int\frac{d^{4}k}{(2\pi)^{4}}
\frac{1}{\pslash-\kslash -m}\frac{1}{k^{2}-m^{2}}f_{0}(k^{2})\nonumber\\
&&=g_{1}g(2\pi)^{4}\delta^{(4)}(p^{\prime}-p)\int\frac{d^{4}k}{(2\pi)^{4}}\frac{1}{k^{2}-m^{2}}f_{0}(k^{2})\\
&&\times\left\{\frac{1}{2}\left[\frac{1}{\pslash-\kslash -m}+\frac{1}{-\pslash-\kslash -m}\right]+\frac{1}{2}\left[\frac{1}{\pslash-\kslash -m}-\frac{1}{-\pslash-\kslash -m}\right]\right\}.\nonumber
\end{eqnarray}
The first term in this expression, which is invariant under $p\rightarrow -p$, depends only on $p^{2}$ and contributes to the mass term, while the second term, which is odd under $p\rightarrow -p$, is proportional to 
$\pslash$ and contributes to the kinetic energy term.

The P- or CP-violating part contained in the mixing of the T-invariant and T-violating interactions gives 
\begin{eqnarray}
&&g_{1}g(2\pi)^{4}\delta^{(4)}(p^{\prime}-p)\int\frac{d^{4}k}{(2\pi)^{4}}\nonumber\\
&&\times i\epsilon\gamma_{5}\left[\left(\frac{1}{-\pslash+\kslash -m}\frac{1}{k^{2}-m^{2}}\right)f_{+}(k)+\frac{1}{\pslash-\kslash -m}\frac{1}{k^{2}-m^{2}}f_{-}(k)\right],
\end{eqnarray}
which is invariant under $p\rightarrow -p$ if one recalls that
$f_{+}(-k)=f_{-}(k)$, and thus contains only $p^{2}$ and contributes to the (parity violating) mass term.

We thus see no asymmetry in positive and negative $p_{0}$, therefore no indication of the particle and anti-particle mass splitting in this setting. This is what we expect, since  our initial Lagrangian is invariant under C, which ensures the equality between masses of the particle and the anti-particle. Still, we see
 that the symmetric appearance of $f_{+}(k)$ and $f_{-}(k)$ in the amplitude, which also ensures the hermiticity, tends to enforce this symmetry between the positive and negative $p_{0}$.

In the present spirit of sample calculations, the next step we need to take is to study a model which violates 
C explicitly in addition to CP, by incorporating chiral gauge interactions, for example. The incorporation of local gauge symmetry into a non-local theory is a rather formidable task.
Instead of analyzing gauge interactions, we later analyze a general class
of free Lagrangians for the fermion, in which  P, C, CP and CPT are formally violated by the present non-local CPT-violating mechanism.

\subsection{ Weinberg's model of CP violation}

We here briefly mention Weinberg's scheme~\cite{weinberg} of possible CP violation in the Standard Model. This model, as it now stands, is not realistic, but it illustrates the possible complications arising from the infrared divergence of the present CPT violation scheme when applied to gauge theory in general. The CP violation in this  model arises from the coupling of the Higgs and the quartet of quarks:
\begin{eqnarray}
{\cal L}_{I}&=&(m_{d}\bar{d}_{R}(x)d_{L}(x)+m_{s}\bar{s}_{R}(x)s_{L}(x))(\lambda_{1}^{\star})^{-1}\varphi_{1}^{0 \star}(x)\nonumber\\
&+&(m_{u}\bar{u}_{R}(x)u_{L}(x)+m_{c}\bar{c}_{R}(x)c_{L}(x))(\lambda_{2})^{-1}\varphi_{2}^{0 }(x)\nonumber\\
&+&[m_{d}\bar{d}_{R}(x)u_{L}(x)
+m_{s}\bar{s}_{R}(x)c_{L}(x)]\int d^{4}y\theta(x^{0}-y^{0})\delta((x-y)^{2}-l^{2})
(\lambda_{1}^{\star})^{-1}\varphi_{1}^{+ \star}(y)\nonumber\\
&-&(m_{u}\bar{u}_{R}(x)d_{L}(x)+m_{c}\bar{c}_{R}(x)s_{L}(x))(\lambda_{2})^{-1}\varphi_{2}^{+}(x) + h.c.,
\end{eqnarray}
where we chose the vanishing Cabbibo angle $\theta_{c}=0$ for simplicity. As an illustration, we inserted the CPT-violating factor into the third term only; we actually need to insert the same factor into the first term to preserve $SU(2)\times U(1)$, but it leads to an infrared divergence arising from the vacuum value of $\varphi_{1}^{0 \star}$, since the vacuum value carries the vanishing momentum. This is an illustration of the possible complications when one attempts to incorporate the present CPT violation into realistic gauge models.

In this model,  CP violation is manifested by the complex 
\begin{eqnarray}
D_{12}(x-y)=\langle T (\lambda_{1}^{\star})^{-1}\varphi_{1}^{+ \star}(x)
(\lambda_{2})^{-1}\varphi_{2}^{+}(y)\rangle.
\end{eqnarray}
If we expand
\begin{eqnarray}
\varphi_{k}^{+}(x)=\sum_{l=1}^{3}a_{kl}\phi_{l}^{+}(x), \ \ \ \ k=1,2
\end{eqnarray}
with the mass eigenstates $\phi_{l}^{+}(x)$, we have
\begin{eqnarray}
D_{12}(x-y)=\sum_{l=1}^{3}(\lambda_{1}^{\star}\lambda_{2})^{-1}a_{1l}^{\star}a_{2l}\langle T \phi_{l}^{+ \star}(x)\phi_{l}^{+}(y)\rangle.
\end{eqnarray}
By performing some detailed analysis, one can confirm that the final conclusion is essentially the same as the above simple model in (4.9).\footnote{After completing our manuscript, we received a related analysis of the
mass splitting induced by Lorentz-invariant CPT violation \cite{dolgov}.}

\section{ General Lagrangians for free fermion}

We study here a general class of Lagangians for a free fermion, which are constructed so that all the  symmetries, P, C, CP and CPT, are formally violated by using the present non-local CPT violation. We also impose formal hermiticity on these Lagrangians. We define two different non-local deformations of the Lagrangian. The first one is the smooth deformation, starting with the conventional CPT-even local theory, with the non-local factor inserted.  This analysis is rather close to  the sample calculations in the preceding 
section, where the CPT violation was studied 
as higher order effects. The second one is a non-smooth deformation, by incorporating terms which are not allowed in the conventional CPT-even local field theory.   

\subsection{CPT violation in the free-quadratic Lagrangian (smooth non-local deformation)}

We examine the Lagrangian
\begin{eqnarray}
{\cal L}&=&\bar{\psi}_{L}(x)i\gamma^{\mu}\partial_{\mu}\psi_{L}(x)+\bar{\psi}_{R}(x)i\gamma^{\mu}\partial_{\mu}\psi_{R}(x)
\nonumber\\
&+&\eta[\bar{\psi}_{R}(x)i\gamma^{\mu}\partial_{\mu}\tilde{F}\psi_{R}(x)+\tilde{F}\bar{\psi}_{R}(x)i\gamma^{\mu}\partial_{\mu}\psi_{R}(x)]
\nonumber\\
&-& m\bar{\psi}(x)\psi(x)+ i\epsilon[\bar{\psi}(x)\gamma_{5}\tilde{F}\psi(x)+\tilde{F}\bar{\psi}(x)\gamma_{5}\psi(x) ],
\end{eqnarray}
where $\eta$ and $\epsilon$ are infinitesimal real parameters.
Here we defined
\begin{eqnarray}
&&\psi_{R,L}(x)=\frac{1}{2}(1\pm \gamma_{5})\psi(x),\nonumber\\
&&\tilde{F}\psi_{R}(x)\equiv \int d^{4}y\theta(x^{0}-y^{0})\delta((x-y)^{2}-l^{2})\psi_{R}(y),\nonumber\\
&&\tilde{F}\psi(x)\equiv \int d^{4}y\theta(x^{0}-y^{0})\delta((x-y)^{2}-l^{2})\psi(y).
\end{eqnarray}
This Lagrangian is formally hermitian and all its terms, when one removes the non-local factor, are allowed and 
CPT-even in the conventional local field theory. 

The first two terms of the Lagrangian in (5.1) preserve all the symmetries. The terms with $\eta$ in the Lagrangian break P, C, T and CPT due to the CPT and T breaking
non-local factor. The terms with $\epsilon$ in the Lagrangian  break P, T, CPT. 
Those breakings are, however, all formal.   
Therefore, we have to examine these apparent symmetry breakings in more detail. 
 
We first note that 
\begin{eqnarray}
\partial_{\mu}\tilde{F}\psi_{R}(x)&=&\int d^{4}y\frac{\partial}{\partial x^{\mu}}\left[\theta(x^{0}-y^{0})\delta((x-y)^{2}-l^{2})\right]\psi_{R}(y)\nonumber\\
&=&-\int d^{4}y\frac{\partial}{\partial y^{\mu}}[\theta(x^{0}-y^{0})\delta((x-y)^{2}-l^{2})]\psi_{R}(y)\nonumber\\
&=&\int d^{4}y\left[\theta(x^{0}-y^{0})\delta((x-y)^{2}-l^{2})\right]\frac{\partial}{\partial y^{\mu}}\psi_{R}(y)\nonumber\\
&=&\tilde{F}\partial_{\mu}\psi_{R}(x)
\end{eqnarray}
after partial integration. We thus have in the action:
\begin{eqnarray}
&&\int d^{4}x\left\{\bar{\psi}_{R}(x)i\gamma^{\mu}\partial_{\mu}\tilde{F}\psi_{R}(x)+\tilde{F}\bar{\psi}_{R}(x)i\gamma^{\mu}\partial_{\mu}\psi_{R}(x)\right\}\nonumber\\
&&=\int d^{4}x\left\{\bar{\psi}_{R}(x)i\gamma^{\mu}\tilde{F}\partial_{\mu}\psi_{R}(x)+\tilde{F}\bar{\psi}_{R}(x)i\gamma^{\mu}\partial_{\mu}\psi_{R}(x)\right\}\nonumber\\
&&=\int d^{4}xd^{4}y\Big\{\bar{\psi}_{R}(x)i\gamma^{\mu}\theta(x^{0}-y^{0})\delta((x-y)^{2}-l^{2})\partial_{\mu}\psi_{R}(y)\nonumber\\
&&+\theta(x^{0}-y^{0})\delta((x-y)^{2}-l^{2})\bar{\psi}_{R}(y)i\gamma^{\mu}\partial_{\mu}\psi_{R}(x)\Big\}\nonumber\\
&&=\int d^{4}xd^{4}y\Big\{\bar{\psi}_{R}(y)i\gamma^{\mu}\theta(y^{0}-x^{0})\delta((y-x)^{2}-l^{2})\partial_{\mu}\psi_{R}(x)\nonumber\\
&&+\theta(x^{0}-y^{0})\delta((x-y)^{2}-l^{2})\bar{\psi}_{R}(y)i\gamma^{\mu}\partial_{\mu}\psi_{R}(x)\Big\}\nonumber\\
&&=\int d^{4}xd^{4}y\delta((x-y)^{2}-l^{2})\bar{\psi}_{R}(y)i\gamma^{\mu}\partial_{\mu}\psi_{R},
\end{eqnarray}
where we interchanged the integration variables $x\leftrightarrow y$ in the 5th line.
This final expression shows that the T-violating effect caused by $\theta(x^{0}-y^{0})$ disappears in the action and CPT invariance is restored.

Similarly, one can show that
\begin{eqnarray}
&&\int d^{4}x\left[i\epsilon\bar{\psi}(x)\gamma_{5}\tilde{F}\psi(x)+i\epsilon\tilde{F}\bar{\psi}(x)\gamma_{5}\psi(x) \right]\nonumber\\
&&=\int d^{4}xd^{4}y\delta((x-y)^{2}-l^{2})[i\epsilon\bar{\psi}(x)\gamma_{5}\psi(y)],
\end{eqnarray}
namely, the T-violating effect caused by $\theta(x^{0}-y^{0})$ disappears in the action and CPT invariance is restored.

This analysis shows that the present non-local CPT violation does not spoil the CPT invariance of the free fermion action if one starts with the conventional CPT-even local theory and deforms it to be non-local. The formal hermiticity of the action is crucial in this analysis. Consequently, no splitting between the masses and the widths of the particle and anti-particle appears. We have already seen in the sample calculations that hermiticity tends to mask CPT breaking effects. We conjecture that the CPT breaking in the free part of the fermion is not induced either directly or as a result of perturbation, as long as one starts with the CPT-even local theory and deforms it by a non-local factor.

We emphasize that this disappearance of CPT violation by the requirement of hermiticity does not take place in the cubic or higher order interaction terms in the Lagrangian. The CPT or T violation implemented by the present non-local scheme is genuine, as was explicitly demonstrated in Section 3.

\subsection{CPT violation in the free-quadratic Lagrangian (non-smooth non-local deformation)}

In the present non-local formulation, we have a new possibility which is absent in a smooth extension of the CPT-even local field theory. The CPT-odd term $i\mu\bar{\psi}(x)\psi(y)$ (to be precise, $i\mu\bar{\psi}(x)\psi(x)$), 
with a real $\mu$, does not appear in the local Lagrangian since it cancels with its hermitian conjugate. But in the present non-local theory one can consider the hermitian combination
\begin{eqnarray}
&&\int d^{4}x\left[i\mu\bar{\psi}(x)\tilde{F}\psi(x)-i\mu\tilde{F}\bar{\psi}(x)\psi(x) \right]\\
&&=\int d^{4}xd^{4}y[\theta(x^{0}-y^{0})-\theta(y^{0}-x^{0})]\delta((x-y)^{2}-l^{2})[i\mu\bar{\psi}(x)\psi(y)],\nonumber
\end{eqnarray}
which is non-vanishing. Under CPT, we have $i\mu\bar{\psi}(x)\psi(y)\rightarrow-i\mu\bar{\psi}(-y)\psi(-x)$. By performing the change of integration variables $-x\rightarrow y$ and $-y \rightarrow x$, this combination  is confirmed to be CPT=$-1$. 
In fact, we have the following transformation properties of the operator part:
\begin{eqnarray}
&& {\rm C}:\ i\mu\bar{\psi}(x)\psi(y)\rightarrow i\mu\bar{\psi}(y)\psi(x),\nonumber\\
&& {\rm P}:\ i\mu\bar{\psi}(x^{0},\vec{x})\psi(y^{0},\vec{y})\rightarrow i\mu\bar{\psi}(x^{0},-\vec{x})\psi(y^{0},-\vec{y}),\nonumber\\
&& {\rm T}:\ i\mu\bar{\psi}(x^{0},\vec{x})\psi(y^{0},\vec{y})\rightarrow -i\mu\bar{\psi}(-x^{0},\vec{x})\psi(-y^{0},\vec{y}),
\end{eqnarray}
and thus the overall transformation property is
C=$-1$, P=$1$,  T=$1$. Namely, C=CP=CPT=$-1$.

It is thus interesting to examine a new action,
\begin{eqnarray}
S&=&\int d^{4}x\Big\{\bar{\psi}(x)i\gamma^{\mu}\partial_{\mu}\psi(x)
 - m\bar{\psi}(x)\psi(x)\\
 && -\int d^{4}y\left[\theta(x^{0}-y^{0})-\theta(y^{0}-x^{0})\right]\delta((x-y)^{2}-l^{2})\left[i\mu\bar{\psi}(x)\psi(y)\right]\Big\},\nonumber 
\end{eqnarray}
which is Lorentz-invariant and hermitian. For the real parameter $\mu$, the third term has $C=CP=CPT=-1$ and no symmetry to ensure the equality of particle and anti-particle masses. 

The Dirac equation is replaced by
\begin{eqnarray}
i\gamma^{\mu}\partial_{\mu}\psi(x)&=&m\psi(x)\\
&+&i\mu\int d^{4}y\left[\theta(x^{0}-y^{0})-\theta(y^{0}-x^{0})\right]\delta((x-y)^{2}-l^{2})\psi(y).\nonumber
\end{eqnarray}
By inserting an ansatz for the possible solution:
\begin{eqnarray}
\psi(x)=e^{-ipx}U(p),
\end{eqnarray}
we have  
\begin{eqnarray}
\pslash U(p)&=&mU(p)\nonumber\\
&+&i\mu\int d^{4}y\left[\theta(x^{0}-y^{0})-\theta(y^{0}-x^{0})\right]\delta((x-y)^{2}-l^{2})e^{-ip(y-x)}U(p)\nonumber\\
&=&mU(p)
+i\mu\left[f_{+}(p)-f_{-}(p)\right]U(p),
\end{eqnarray}
where $f_{\pm}(p)$ is the Lorentz-invariant form factor defined in Appendix A. The propagator is defined by 
\begin{eqnarray}
\int d^{4}x e^{ipx}\langle T^{\star}\psi(x)\bar{\psi}(0)\rangle=\frac{i}{\pslash-m +i\epsilon-i\mu\left[f_{+}(p)-f_{-}(p)\right]},
\end{eqnarray}
which is Lorentz covariant. Note that we use $T^{\star}$-product for the path integral on the basis of Schwinger's action principle, which is based on the equation of motion (5.9) with a source term added,
\begin{eqnarray}
\langle 0,+\infty|0,-\infty\rangle_{J}=\int{\cal D}\bar{\psi}{\cal D}\psi\exp\left\{iS+i\int d^{4}x {\cal L}_{J}\right\},
\end{eqnarray}
where the action $S$ is given in (5.8) and the source term is ${\cal L}_{J}=\bar{\psi}(x)\eta(x)+\bar{\eta}(x)\psi(x)$. The $T^{\star}$-product
is quite different from the canonical $T$-product in the present non-local theory; in fact, the canonical quantization is not defined in the present theory.

For space-like $p$, the extra term with $\mu$ in the denominator of the propagator (5.12) vanishes, since $f_{+}(p)=f_{-}(p)$ for $p=(0,\vec{p})$, as is shown in Appendix A, eq. (A.3). Thus, the propagator has poles only at time-like momentum, and in this sense the present hermitian action (5.8) does not allow a tachyon.
By assuming a time-like $p$, we go to the frame where $\vec{p}=0$.
Then the eigenvalue equation is given by 
\begin{eqnarray}
p_{0}\gamma^{0}&=&m + i\mu [f_{+}(p_{0})-f_{-}(p_{0})],
\end{eqnarray}
namely,
\begin{eqnarray}
p_{0}\gamma^{0}&=&m - 4\pi \mu\int_{0}^{\infty}dz\frac{z^{2}\sin [ p_{0}\sqrt{z^{2}+l^{2}}]}{\sqrt{z^{2}+l^{2}}},
\end{eqnarray}
where we used  the explicit formula (A.4),
\begin{eqnarray}
f_{\pm}(p_{0})
=2\pi \int_{0}^{\infty}dz\frac{z^{2}e^{\pm ip_{0}\sqrt{z^{2}+l^{2}}}}{\sqrt{z^{2}+l^{2}}}.
\end{eqnarray}
The solution $p_{0}$ of eq. (5.15) determines the possible mass eigenvalues. 

This eigenvalue equation becomes, under $p_{0}\rightarrow -p_{0}$:
\begin{eqnarray}
-p_{0}\gamma^{0}
&=&m + 4\pi\mu\int_{0}^{\infty}dz\frac{z^{2}\sin [ p_{0}\sqrt{z^{2}+l^{2}}]}{\sqrt{z^{2}+l^{2}}}.
\end{eqnarray}
By sandwiching this equation by $\gamma_{5}$, which is regarded as CPT operation, we have:
\begin{eqnarray}
-p_{0}\gamma^{0}
&=&-m - 4\pi\mu\int_{0}^{\infty}dz\frac{z^{2}\sin [ p_{0}\sqrt{z^{2}+l^{2}}]}{\sqrt{z^{2}+l^{2}}},
\end{eqnarray}
namely,
\begin{eqnarray}
p_{0}\gamma^{0}
&=&m + 4\pi\mu\int_{0}^{\infty}dz\frac{z^{2}\sin [ p_{0}\sqrt{z^{2}+l^{2}}]}{\sqrt{z^{2}+l^{2}}},
\end{eqnarray}
which is not identical to the original equation in (5.15).
In other words, if $p_{0}$ is the solution of the original equation, $-p_{0}$ cannot be the solution of the original equation except for $\mu=0$. The last term in the Lagrangian (5.8) with C=CP=CPT=$-1$ splits the particle and anti-particle masses. As a very crude estimate of the mass splitting, one may assume $\mu\ll m$ and solve these equations iteratively. If the particle mass for (5.15) is chosen at 
\begin{eqnarray}
p_{0}\simeq m - 4\pi \mu\int_{0}^{\infty}dz\frac{z^{2}\sin [ m \sqrt{z^{2}+l^{2}}]}{\sqrt{z^{2}+l^{2}}},
\end{eqnarray}
then the anti-particle mass for (5.19) is estimated at 
\begin{eqnarray}
p_{0}\simeq m + 4\pi \mu\int_{0}^{\infty}dz\frac{z^{2}\sin [ m \sqrt{z^{2}+l^{2}}]}{\sqrt{z^{2}+l^{2}}}.
\end{eqnarray}
This simple Lagrangian model in (5.8) may provide a useful theoretical  laboratory when one investigates  Lorentz-invariant CPT violation in the future, for example, in connection with the neutrino mass~\cite{murayama}.

Further detailed analyses of this  Lagrangian model were reported elsewhere~\cite{CFT}. 

\subsection{Noether's theorem}

It is interesting to see how Noether's theorem for the global fermion number symmetry in the non-local action (5.8) is realized.
The action is invariant under the fermion number phase transformation,
\begin{eqnarray}
\psi(x)\rightarrow e^{i\alpha}\psi(x), \ \ \ \ \bar{\psi}(x)\rightarrow \bar{\psi}(x)e^{-i\alpha}
\end{eqnarray}
with a real constant $\alpha$.

In the path integral formulation of Noether's theorem or Ward--Takahashi identities~\cite{fujikawa3}, we define
\begin{eqnarray}
\psi^{\prime}(x)= e^{i\alpha(x)}\psi(x), \ \ \ \ \bar{\psi}^{\prime}(x)=\bar{\psi}(x)e^{-i\alpha(x)}
\end{eqnarray}
and start with the identity (we neglect the source term, as it is not essential in the present analysis):
\begin{eqnarray}
\int{\cal D}\bar{\psi}{\cal D}\psi\exp\{iS(\bar{\psi},\psi)\}=\int{\cal D}\bar{\psi}^{\prime}{\cal D}\psi^{\prime}\exp\{iS(\bar{\psi}^{\prime}, \psi^{\prime})\},
\end{eqnarray}
which means that the naming of integration variables does not change the integral itself.
In the absence of anomaly, as is the present case, we have~\cite{fujikawa3}
\begin{eqnarray}
{\cal D}\bar{\psi}{\cal D}\psi={\cal D}\bar{\psi}^{\prime}{\cal D}\psi^{\prime}
\end{eqnarray}
and thus the above identity is written as 
\begin{eqnarray}
\int{\cal D}\bar{\psi}{\cal D}\psi\exp\{iS(\bar{\psi},\psi)\}=\int{\cal D}\bar{\psi}{\cal D}\psi\exp\{iS(\bar{\psi}^{\prime}, \psi^{\prime})\},
\end{eqnarray}
which is the basis of field theoretical indentities.
We obtain the Ward--Takahashi-type identities when expanding $S(\bar{\psi}^{\prime}, \psi^{\prime})$ in powers of the infinitesimal $\alpha(x)$.
In the lowest order of $\alpha(x)$, we have 
\begin{eqnarray}
S(\bar{\psi}^{\prime}, \psi^{\prime})&=&S(\bar{\psi}, \psi)+\int d^{4}x\alpha(x)\partial_{\mu}[\bar{\psi}(x)\gamma^{\mu}\psi(x)]\\
&&-
\int d^{4}x\alpha(x)\mu\int d^{4}y[\theta(x^{0}-y^{0})-\theta(y^{0}-x^{0})]\delta((x-y)^{2}-l^{2})[\bar{\psi}(x)\psi(y)]\nonumber\\ 
&&+\int d^{4}y\alpha(y) \mu \int d^{4}x[\theta(x^{0}-y^{0})-\theta(y^{0}-x^{0})]\delta((x-y)^{2}-l^{2})[\bar{\psi}(x)\psi(y)] \nonumber
\end{eqnarray}
after partial integration. When one uses this relation in the identity
(5.26) and when one expands the factor in the exponential to the linear order of $\alpha(x)$, one obtains: 
\begin{eqnarray}
&&\Big\langle\int d^{4}x\alpha(x)\partial_{\mu}T^{\star}[\bar{\psi}(x)\gamma^{\mu}\psi(x)]\nonumber\\
&&-
\int d^{4}x\alpha(x)\mu\int d^{4}y[\theta(x^{0}-y^{0})-\theta(y^{0}-x^{0})]\delta((x-y)^{2}-l^{2})T^{\star}[\bar{\psi}(x)\psi(y)]\nonumber\\ 
&&+\int d^{4}y\alpha(y) \mu \int d^{4}x[\theta(x^{0}-y^{0})-\theta(y^{0}-x^{0})]\delta((x-y)^{2}-l^{2})T^{\star}[\bar{\psi}(x)\psi(y)]\Big\rangle \nonumber\\
&&=0.
\end{eqnarray}
When one applies the functional derivative $\frac{\delta}{\delta\alpha(z)}$ to this relation, one obtains
\begin{eqnarray}
\partial_{\mu}\langle T^{\star}\bar{\psi}(z)\gamma^{\mu}\psi(z)\rangle
&=&
\Big\langle\mu\int d^{4}y[\theta(z^{0}-y^{0})-\theta(y^{0}-z^{0})]\delta((z-y)^{2}-l^{2})T^{\star}[\bar{\psi}(z)\psi(y)]\Big\rangle\nonumber\\ 
&&-\Big\langle \mu \int d^{4}x[\theta(x^{0}-z^{0})-\theta(z^{0}-x^{0})]\delta((x-z)^{2}-l^{2})T^{\star}[\bar{\psi}(x)\psi(z)]\Big\rangle \nonumber\\
&=&
\Big\langle\mu\int d^{4}y[\theta(z^{0}-y^{0})-\theta(y^{0}-z^{0})]\delta((z-y)^{2}-l^{2})T^{\star}[\bar{\psi}(z)\psi(y)]\Big\rangle\nonumber\\ 
&&-\Big\langle \mu \int d^{4}y[\theta(y^{0}-z^{0})-\theta(z^{0}-y^{0})]\delta((z-y)^{2}-l^{2})T^{\star}[\bar{\psi}(y)\psi(z)]\Big\rangle.\nonumber\\ 
\end{eqnarray}
The relations (5.28) and (5.29) are the general identities for the fermion number symmetry in the present non-local theory, which give rise to general Ward--Takahashi identities when inserted into Green's functions.


The relation (5.29) is the statement of the fermion number conservation (Noether's theorem) in the present non-local theory defined by (5.8) and the non-local equation of motion in (5.9). We avoid the use of the term "charge" for this symmetry since we have no gauge symmetry and thus no electromagnetic field to detect the charge. A general analysis of electromagnetic interactions in this model will be given elsewhere~\cite{CFT3}, where the relation corresponding to (5.29) is shown to give rise to the electric charge conservation\footnote{An opposite result has been reported in \cite{dolgov}.}.

\section{Discussion and conclusion}

The results of our analysis may be summarized as follows.

1. As concerns the formal aspects of quantizing the Lagrangian  non-local in time, there is no definite method which is satisfactory in all respects. The path integral on the basis of Schwinger's action principle adopted in the present paper is one of the possible working hypotheses. The final outcomes  of the formulation in  lower order perturbation theory are reasonable, which suggests that the path integral scheme is sensible.  

2. The non-local implementation of CPT and T violation was shown to produce the expected T violation in decay and production processes
in Section 3. This suggests that the induced dipole moment, decay or scattering amplitudes, in general, can contain the CPT- or T-violation effects.  

3. Although we have not completed the analysis of all the possible induced CPT violations, our analysis so far strongly indicates that the perturbative treatment of the present non-local CPT violation, which is implemented in the CPT-even local Lagrangian, does not spoil the CPT invariance  of the free part of the Lagrangian.
A treatment of chiral gauge theory in the present scheme, which breaks
C invariance, is in general a difficult but important future task to
complete the analysis of the induced CPT violation in the present
scheme. However, relaxing the quadratic CPT-violating term in the
Lagrangian used in  the present work as in (5.8),  the question remains
as of which symmetry is  responsible for the equality of the masses of
partice and  antiparticle. By taking CPT violation to be due to only
interaction  (when C and CP are also violated),  two of the present authors  (MC and AT) will analyze elsewhere~ \cite{CT} whether  in such a case the equality of masses is maintained.
A further clarification of the basic mechanism which may ensure
equal masses to the particle and anti-particle  in the absence of C, CP
and CPT symmetries is an interesting remaining issue.

4. We have found an interesting simple Lagrangian model (5.8) which produces the splitting of particle and anti-particle masses by the present Lorentz-invariant CPT violation.
The detailed analysis of this  Lagrangian model was reported elsewhere~\cite{CFT}, and its application to the neutrino anti-neutrino mass splitting in the Standard Model was also presented and the unique nature of the neutrino mass was emphasized~\cite{CFT2}. 

5. As for the mathematical subtlety of the present formulation, in particular, in the analyses in Section 3 and Section 5, 
we treated the form factor 
\begin{eqnarray}
f_{\pm}(k^{0})
&=&2\pi \int_{0}^{\infty}dz\frac{z^{2}e^{\pm ik^{0}\sqrt{z^{2}+l^{2}}}}{\sqrt{z^{2}+l^{2}}}
\end{eqnarray}
as a well-defined function instead of a distribution\footnote{
 It is possible to assign a finite value to the last term in eqs.(5.20) and (5.21) for $p_{0}\neq 0$ by using the formal relation
\begin{eqnarray}
\int_{0}^{\infty}dz\frac{z^{2}\sin [
p_{0}\sqrt{z^{2}+l^{2}}]}{\sqrt{z^{2}+l^{2}}}&=&-\frac{\partial^{2}}{\partial p_{0}^{2}}\int_{0}^{\infty}dz\frac{z^{2}\sin [p_{0}\sqrt{z^{2}+l^{2}}]}{[z^{2}+l^{2}]^{3/2}}.\nonumber
\end{eqnarray}}. As is explained in Appendix A, this is mathematically related to a precise understanding of the two-point Wightman function for a free scalar field at time-like separation.

6. One may attempt to modify the model Lagrangian in (1.1) to be more realistic, for example, 
\begin{eqnarray}
{\cal L}&=&\bar{\psi}(x)[i\gamma^{\mu}\partial_{\mu}-g(\phi(x)+i\gamma_{5}\varphi(x))]\psi(x)\nonumber\\
&& +
\frac{1}{2}\partial_{\mu}\phi(x)\partial^{\mu}\phi(x)+\frac{1}{2}\partial_{\mu}\varphi(x)\partial^{\mu}\varphi(x)
-V(\phi,\varphi)
\nonumber\\
&&+ g_{1}\bar{\psi}(x)\int d^{4}y\theta(x^{0}-y^{0})\delta((x-y)^{2}-l^{2})[\phi(y)+i\gamma_{5}\varphi(y)]\psi(x),
\end{eqnarray}
with
\begin{eqnarray}
V(\phi,\varphi)&=&\frac{1}{4}{\rm Tr}\left\{\frac{\mu^{2}}{2}(\phi+i\gamma_{5}\varphi)(\phi-i\gamma_{5}\varphi)+\frac{\lambda}{4!}
\left[(\phi+i\gamma_{5}\varphi)(\phi-i\gamma_{5}\varphi)\right]^{2}\right\}\nonumber\\
&=&\frac{\mu^{2}}{2}(\phi^{2}+\varphi^{2})+\frac{\lambda}{4!}
(\phi^{2}+\varphi^{2})^{2},
\end{eqnarray}
where ${\rm Tr}$ stands for the trace over the Dirac matrix.
This Lagrangian is invariant under the global chiral transformation
\begin{eqnarray}
&&\psi(x)\rightarrow U(\alpha)\psi(x),\nonumber\\
&&(\phi+i\gamma_{5}\varphi)\rightarrow U(\alpha)(\phi+i\gamma_{5}\varphi)U(\alpha)^{\dagger},
\end{eqnarray}
where $U(\alpha)=\exp[i\gamma_{5}\alpha]$ with a space-time independent parameter $\alpha$. One may attempt to generate the fermion mass by a Higgs-type mechanism,
\begin{eqnarray}
\phi(x)\rightarrow \phi(x)+v,
\end{eqnarray}
with $\varphi(x)$ standing for the massless Nambu--Goldstone field.
In this class of analysis, the infrared divergent form factor 
leads to the problem with the constant $v$ and the massless $\varphi(x)$. The treatment of the spontaneous symmetry breaking needs to be understood better in the more realistic applications of the present CPT violation mechanism even in the context of non-gauge theory. 

\paragraph{ Note added in proof.}

Our form factor (1.5) (see also (3.6), (4.4), (4.7), (6.1), (A.1) and
(A.6)) is related to the two-point Wightman function for a real free
scalar field when the coordinates and momenta are interchanged. The T-
or CPT-violation by an apparently real-looking form factor in our model
is related to the fact that the form factor develops an imaginary part
for the time-like momentum, which corresponds to a cut in the Wightman
function for a time-like separation. The referee raised an interesting question, whether the CPT violation in our model is related to the negative norm states introduced by the non-local form factor. This issue certainly deserves further investigation, but the negative norm by itself does not induce CPT violation, as is illustrated by the Faddeev--Popov hermitian Lagrangian for the Landau gauge in QED, ${\cal L} = B\partial^{\mu}A_{\mu} - i\xi\partial^{\mu}\partial_{\mu}\eta$, with real Faddeev--Popov ghost $\eta$ and anti-ghost $\xi$ which induces an indefinite norm but does not violate CPT: under anti-unitary CPT,
$A_{\mu}(x^{0},\vec{x})\rightarrow -A_{\mu}(-x^{0},-\vec{x})$,
$B(x^{0},\vec{x})\rightarrow B(-x^{0},-\vec{x})$,
$\xi(x^{0},\vec{x})\rightarrow \xi(-x^{0},-\vec{x})$, and
$\eta(x^{0},\vec{x})\rightarrow -\eta(-x^{0},-\vec{x})$.

\section*{Acknowledgements}

The support of the Academy
of Finland under the Projects No. 136539 and 140886 is gratefully 
 acknowledged.

\appendix

\section{Appendix: Form factor responsible for CPT violation}

It has been noted in the body of the paper that the present way of incorporating CPT violation is realized by an extra form factor in the interaction vertex (see eq. (1.5)). It was also noted that this form factor is infrared divergent. We here analyze some details of the infrared structure of the CPT-violating form factor. We define two Lorentz-invariant form factors,
\begin{eqnarray}
&&f_{\pm}(k)=\int d^{4}z_{1}
e^{\pm ikz_{1}}\theta(z_{1}^{0})\delta((z_{1})^{2}-l^{2}),
\end{eqnarray}
which are  inequivalent for time-like $k$ due to the time ordering factor $\theta(z_{1}^{0})$.
These form factors contain quadratic divergences at $k\rightarrow 0$. To be specific,
\begin{eqnarray}
f_{\pm}(0)&=&\int d^{4}z_{1}\theta(z_{1}^{0})\delta((z_{1})^{2}-l^{2})\nonumber\\
&=&\int d^{3}z_{1}\frac{1}{2\sqrt{\vec{z}_{1}^{2}+l^{2}}}\nonumber\\
&\sim& L^{2} \rightarrow \infty,
\end{eqnarray}
with $L$ the size of the space. For space-like momentum, one may choose a suitable Lorentz frame such that $k^{0}=0$, and then
\begin{eqnarray}
f_{\pm}(\vec{k})
&=&\int d^{3}z_{1}\frac{e^{\mp i\vec{k}\vec{z}_{1}}}{2\sqrt{\vec{z}_{1}^{2}+l^{2}}}\nonumber\\
&=&2\pi\int d|z_{1}||z_{1}|^{2}\frac{\sin|k||z_{1}|}{|k||z_{1}|\sqrt{|z_{1}|^{2}+l^{2}}}\nonumber\\ 
&=&\frac{2\pi}{|k|^{2}}\int_{0}^{\infty} dz z \frac{\sin z}{\sqrt{z^{2}+(|k|l)^{2}}},
\end{eqnarray}
which is analogous to the Fourier transform of the Coulomb potential and diverges quadratically at $\vec{k}\rightarrow 0$. For the time-like momentum, one may choose $\vec{k}=0$, and 
\begin{eqnarray}
f_{\pm}(k^{0})
&=&\int d^{3}z_{1}\frac{e^{\pm ik^{0}\sqrt{\vec{z}_{1}^{2}+l^{2}}}}{2\sqrt{\vec{z}_{1}^{2}+l^{2}}}\nonumber\\
&=&2\pi \int_{0}^{\infty}dz\frac{z^{2}e^{\pm ik^{0}\sqrt{z^{2}+l^{2}}}}{\sqrt{z^{2}+l^{2}}},
\end{eqnarray}
or, if one assumes $k^{0}>0$,
\begin{eqnarray}
f_{\pm}(k^{0})
&=&\frac{2\pi}{(k^{0})^{2}} \int_{0}^{\infty}dz\frac{z^{2}e^{\pm i\sqrt{z^{2}+(k^{0}l)^{2}}}}{\sqrt{z^{2}+(k^{0}l)^{2}}},
\end{eqnarray}
which again diverges quadratically for $k^{0}\rightarrow 0$.

In passing, we recall the two-point Wightman function for a free scalar field:
\begin{eqnarray}
\langle 0|\phi(x)\phi(y)|0\rangle
&=&\int\frac{d^{4}k}{(2\pi)^{4}}e^{-ik(x-y)}2\pi\delta(k^{2}-m^{2})\theta(k^{0}).
\end{eqnarray}
It is interesting that this expression, when one exchanges coordinates and momenta, precisely gives rise to the CPT-violating form factor in (A.1).
The quadratic infrared divergence corresponds to the quadratic divergence of the Wightman function at short distances $x-y \rightarrow 0$. This analogy shows that our form factor is well-defined for $k\neq 0$ at least as a distribution. The precise behavior of the Wightman function at time-like separation is important in order to understand our form factor for time-like momentum.

This quadratic infrared divergence does not lead to an infrared  divergence in the calculations of the self-energy corrections to fermions (except for the case of Weinberg's model of CP violation, where it introduces certain complications), but it gives rise to a divergence in the decay amplitudes in Section 3 at $k\rightarrow 0$.

For the case of the CPT-violating factor $\theta(z_{1}^{0})\theta((z_{1})^{2}-l^{2})$, 
which is the original choice in \cite{CFT}, one has 
\begin{eqnarray}
f_{\pm}(k)=\int d^{4}z_{1}
e^{\pm ikz_{1}}\theta(z_{1}^{0})\theta((z_{1})^{2}-l^{2})
\end{eqnarray}
and thus 
\begin{eqnarray}
f_{\pm}(0)&=&\int d^{4}z_{1}\theta(z_{1}^{0})\theta((z_{1})^{2}-l^{2})\nonumber\\
&=&\int_{l}^{\infty}dz_{1}^{0}\int_{z_{1}^{0}\geq \sqrt{\vec{z}_{1}^{2}+l^{2}}}d^{3}z_{1}\nonumber\\
&=&\int_{l}^{\infty}dz_{1}^{0}\frac{4\pi}{3}\left[\sqrt{(z_{1}^{0})^{2}-l^{2}}\right]^{3},
\end{eqnarray}
which diverges quartically and induces an infrared divergence in the self-energy corrections to fermions. For the space-like momentum, one may choose a suitable Lorentz frame such that $k^{0}=0$. Then
\begin{eqnarray}
f_{\pm}(\vec{k})
&=&\int_{l}^{\infty}dz_{1}^{0}\int_{z_{1}^{0}\geq \sqrt{\vec{z}_{1}^{2}+l^{2}}}e^{\mp i\vec{k}\vec{z}_{1}}d^{3}z_{1}\nonumber\\
&=&\int_{l}^{\infty}dz_{1}^{0}\int_{z_{1}^{0}\geq \sqrt{|\vec{z}_{1}|^{2}+l^{2}}}2\pi\frac{e^{i|\vec{k}||\vec{z}_{1}|}-e^{-i|\vec{k}||\vec{z}_{1}|}}
{i|\vec{k}||\vec{z}_{1}|}|\vec{z}_{1}|^{2}d|\vec{z}_{1}|\nonumber\\
&=&\frac{4\pi}{|\vec{k}|^{4}}\int_{|\vec{k}|l}^{\infty}d\tilde{z}_{1}^{0}\int_{0}^{\sqrt{(\tilde{z}_{1}^{0})^{2}-(|\vec{k}|l)^{2}}}\sin(z)zdz,
\end{eqnarray}
which diverges quartically for $|\vec{k}|\rightarrow 0$. For time-like $k$, one may choose the frame where $\vec{k}=0$. Then one has 
\begin{eqnarray}
f_{\pm}(k^{0})
&=&\int_{l}^{\infty}dz_{1}^{0}e^{\pm ik^{0}z^{0}_{1}}\int_{z_{1}^{0}\geq \sqrt{\vec{z}_{1}^{2}+l^{2}}}d^{3}z_{1}\nonumber\\
&=&\int_{l}^{\infty}dz_{1}^{0}e^{\pm ik^{0}z^{0}_{1}}\frac{4\pi}{3}\left[\sqrt{(z_{1}^{0})^{2}-l^{2}}\right]^{3}\nonumber\\
&=&\frac{4\pi}{3(k^{0})^{4}}\int_{k^{0}l}^{\infty}dze^{\pm iz}\left[\sqrt{z^{2}-(k^{0}l)^{2}}\right]^{3},
\end{eqnarray}
which diverges quartically for $k^{0}\rightarrow 0$.

These infrared divergences are related to the fact that we cannot divide Minkowski space into domains with finite {\em 4-dimensional} volumes in a Lorentz-invariant manner. The best  we can do is to use $\theta(z_{1}^{0})\delta((z_{1})^{2}-l^{2})$ which still has a quadratically divergent volume. 
Note that the local field theory corresponds to the 
choice $\delta^{4}(z)$ and to a constant form factor  $f(k)=1$ for any $k$.

\end{document}